\documentclass[12pt,preprint]{aastex}
\usepackage{emulateapj5}
\def\myputfigure#1#2#3#4#5
{\hskip0.03\textwidth\vskip#5pt
\makebox[0pt]{\hskip#2in
\includegraphics[width=#3\textwidth]{#1}}\vskip#4pt\hfill}
\def \bref{\par \noindent \hangindent=1.5 truecm \hangafter=1}
\def \kms       {\hbox{ km s$^{-1}$}}
\def \se        {\!=\!}
\def \sims      {\sim \!}

\def \spropto   {\! \propto \!}

\def\\{\hfil\break}
\def\spose#1{\hbox to 0pt{#1\hss}}
\def\lta{\mathrel{\spose{\lower 3pt\hbox{$\mathchar"218$}}
     \raise 2.0pt\hbox{$\mathchar"13C$}}}
\def\gta{\mathrel{\spose{\lower 3pt\hbox{$\mathchar"218$}}
     \raise 2.0pt\hbox{$\mathchar"13E$}}}

\begin{document}

\title{Detecting Intracluster Gas Motion in Galaxy Clusters: Mock {\it Astro-E2} Observations} 

\author{Andrew Pawl\altaffilmark{1}, August E. Evrard\altaffilmark{1,2,3},
Renato A. Dupke\altaffilmark{1,2}}
\email{apawl@umich.edu, evrard@umich.edu, rdupke@umich.edu}
\altaffiltext{1}{Michigan Center for Theoretical Physics, University
  of Michigan, Ann Arbor, MI} 
\altaffiltext{2}{Department of Astronomy, University of Michigan, Ann
  Arbor, MI}
\altaffiltext{3}{Department of Physics, University of Michigan, Ann
  Arbor, MI}

\begin{abstract}
We explore the detectability of bulk motions in the X-ray
emitting intracluster medium (ICM) using a catalog of 1,836 mock {\it Astro-E2}
observations of simulated clusters of galaxies.  We generate high
resolution mock spectra for two observing strategies: a four-pointing
mosaic and a single central pointing.  Normalizing to 200 (400) photons in
the iron K$\alpha$ region for the mosaic (central) study, 
we fit Poisson realizations of each
simulated spectrum to a velocity  
broadened isothermal plasma emission model.  We confirm that the
velocity characteristics (mean and dispersion) returned by the spectral fittings
are unbiased measures of the emission-weighted values 
within the observed region, with scatter $\pm 55 \kms$.  The maximum
velocity difference between mosaic element pairs $\Delta v_{\rm max}$ has
$\sims 6\%$ likelihood of being transonic ($\Delta v_{\rm max}
\ge 0.5 c_s$), and the likelihood falls steeply, $p \spropto (\Delta v_{\rm
  max}/c_s)^{-4}$, at high Mach number.  The velocity broadening
parameter $\sigma_v$ from the central pointing fit exceeds the thermal
value in $49\%$ of the cases, 
with again a $\sigma_v^{-4}$ tail at large dispersion. 
We present as case studies the clusters that yield the strongest signal for each 
observing strategy.

\end{abstract}


\section{Introduction}

In a hierarchical model of structure formation, galaxy clusters are
built from the continuous accretion of sub-systems and, therefore,
clusters are expected to frequently display visible signatures of
mergers.  Binary clusters (Jones \& Forman 1984) offer evidence of
precursor systems while clusters with multiple emissivity peaks,
strong isophotal twists and ellipticity variations possess 
morphologies associated with the later 
stages of merger evolution (Mohr, Fabricant \& Geller 1993; Buote \& Tsai 1996).  
Bulk motions and turbulence of the hot intracluster medium (ICM) are
thus expected to be evident in line emission from the 
plasma, in particular from helium-like iron, FeXXV 
(Dupke \& Bregman 2001a,b; Sunyaev et al. 2003; Inogamov \& Sunyaev 2003).   

X-ray satellite missions to date have had spectrometers with 
energy resolutions marginally adequate for the observation of cluster
velocity structure 
(Dupke \& Bregman 2001a,b).  The new microcalorimeter technology
employed in the X-ray Spectrometer (XRS) instrument aboard the
{\it Astro-E2} satellite, however, is designed to have an energy resolution
of 6.5 eV FWHM across its entire frequency range (Furusho et al. 2004),
giving a response contribution to the line broadening of $\sim$1.5 eV 
and finding line centroids to within about 2 eV.
This implies that the instrument will be sensitive to
velocity structure down to $\sim$100 
km/s near the Fe K$\alpha$ line.  Since the infall velocity of
material accreting onto a large cluster can be an order of
magnitude larger than this limit, {\it Astro-E2} will enable unambiguous
observation of gas motion in dynamically active systems.

Cosmological simulations provide a framework for investigating the
statistical properties of the internal gas velocity distribution. 
In addition, access to the complete 3-D structure and time 
evolution of the simulated clusters can, through comparative studies
of observations and simulations, help unfold the 
evolutionary history and dynamical state of observed clusters. 

Until now, no estimate of the frequency with which clusters should
show detectable velocity structure has been made, although Fujita et al. (2005)
have used mock observations of simulated clusters to evaluate the
detectability of turbulence in a specific case study.  In this paper, we
evaluate the fraction of clusters expected to show observable
structure in the Fe K$\alpha$ line by performing mock {\it Astro-E2} observations
of a catalog of simulated clusters.  The simulations model clusters in
a $\Lambda$CDM cosmology with a preheated assumption for the ICM that
matches observed local scaling laws (Bialek, Evrard \& Mohr 2001).  

In \S2, we describe the simulation sample and the strategies we use to
produce mock observations.  Results in the form of statistical
expectations and illustrative examples are presented in \S3.  We
summarize our findings in \S4.

\section{Mock Observations}

\subsection{Preheated Cluster Simulations}

We use a catalog of 68 clusters evolved with P3MSPH (Evrard
1988) under a preheated assumption for the ICM.  The level of entropy
introduced into the initial conditions is tuned to $105.9$~keV~cm$^2$
in order to match the observed scalings of luminosity and ICM mass
with temperature (Bialek et al. 2001).  

The simulations were produced using a multi-step procedure outlined in
Bialek et al. (2001).  The underlying cosmology is a flat, concordance model
with $\Omega_{m} = 0.3$, $\Omega_{\Lambda} = 0.7$, $\Omega_{b} =
0.03$, $\sigma_8 = 1.0$, and $h = 0.7$, where the Hubble constant is defined
as $100h$~km~s$^{-1}$~Mpc$^{-1}$ and $\sigma_8$ is the power spectrum
normalization on $8h^{-1}$ Mpc scales.  Details of the full ensemble
will be presented in Bialek, Evrard \& Mohr (2005).    
One member of the ensemble contained a prominent cold front feature,
generated during a merger by the separation and resultant adiabatic
cooling of the core of an infalling satellite, as discussed by
Bialek, Evrard \& Mohr (2002). 

The cosmological values assumed for the models are in line with {\it WMAP}
allowed parameters, with the exception of the baryon fraction.  
The value $\Omega_b/\Omega_m \se 0.1$ used in the simulations is
lower than the {\it WMAP} value $0.17 \pm 0.01$ (Bennett et al. 2003).
The mass fraction in intracluster gas is expected to be less than the
cosmic baryon fraction due to galaxy formation and energy exchange with
dark matter during mergers (Thomas \& Couchman 1992).  The former
process removes $\sims 15-20\%$ of the baryons from the hot phase in rich
clusters while the latter expels $\sim$10$\%$ of the gas from
the potential (Frenk et al. 1999).  Still, the model gas fractions are
likely to be somewhat low compared
to current observational estimates.  However, in the results presented
here, the baryon fraction affects only the normalization of the ICM
X-ray spectrum.  Our mock spectra are tuned to a fixed line
normalization, which at fixed cluster mass scales $\spropto
(\Omega_b/\Omega_m)^2 t_{\rm exp}$, with $t_{\rm exp}$ the exposure
time.  A higher or lower baryon fraction can thus be absorbed by
appropriate rescaling of the exposure.  

The configuration of each simulation is stored 
at twenty output times, spaced equally in time from the initial
redshift $z_i \se 20.82$ to the present. 
 We employ the final nine outputs of the 68 models in this study,
corresponding to the redshifts 0.540, 0.448, 0.365, 0.290, 0.222, 0.160, 0.102, 
0.049 and 0.0.  
This
yields 612 cluster realizations which we treat, under an ergodic
hypothesis, as statistically independent.  The time separation of $\sims
0.7$~Gyr between outputs typically exceeds the crossing time at
$r_{200}$, justifying the ergodic assumption.  

We further enhance our sample by considering the line of sight (LOS)
velocity structure of each realization along three perpendicular axes.  
Although these projections are linked by dynamics, the 
LOS velocity structure is genuinely independent.  
We thus have three projections of each of the 612 realizations, for a total
of 1,836 exposures.  The members of the simulated sample range in
spectral temperature from 1.5 keV to  
about 8 keV, with cluster masses $M_{200}$ ranging from 0.015-2.4 $\times
10^{15} M_{\odot}$.

The derived spectra and images, along with associated parameters describing the
simulations are publicly available as part of VCE, the {\it Virtual Cluster
  Exploratory}\footnotemark[4]. 
\footnotetext[4]{\url{http://vce.physics.lsa.umich.edu}}

\subsection{Creating the Mock Exposures}
\label{sec:mockex}

We create mock {\it Astro-E2} spectra of these realizations following two
observing programs illustrated in Figure~\ref{fig:tiles}.
The first program involves a search for velocity gradients by
measuring differences in the mean velocity of the Fe K$\alpha$ lines
for multiple pointings of the {\it Astro-E2} instrument. This strategy
allows one to measure both the magnitude and a crude direction of the
velocity gradient. Multiple pointings are required because 
the average angular resolution of the
X-ray telescopes on {\it Astro-E2} is $\sim$1.9$^\prime$ (half-power
diameter\footnotemark[5]) 
\footnotetext[5]{\url{http://heasarc.gsfc.nasa.gov/docs/astroe/prop\_tools/astroe2\_td/}} and the 
detector configuration is roughly a square with 2.9$^\prime$ on the
side. Since less than 60\% of the encircled energy is contained  
within 2$^\prime$, spatially resolved spectroscopy using sub-regions of the detector will be highly contaminated 
with photons from neighboring regions. Therefore, to carry out this observational strategy, we create four
mock spectra of each cluster projection in a 2$\times$2 box centered
on the cluster.  The second program involves searching for
extra-thermal broadening of the Fe K$\alpha$ line using a
single, central pointing.

\vskip-0.1in
\myputfigure{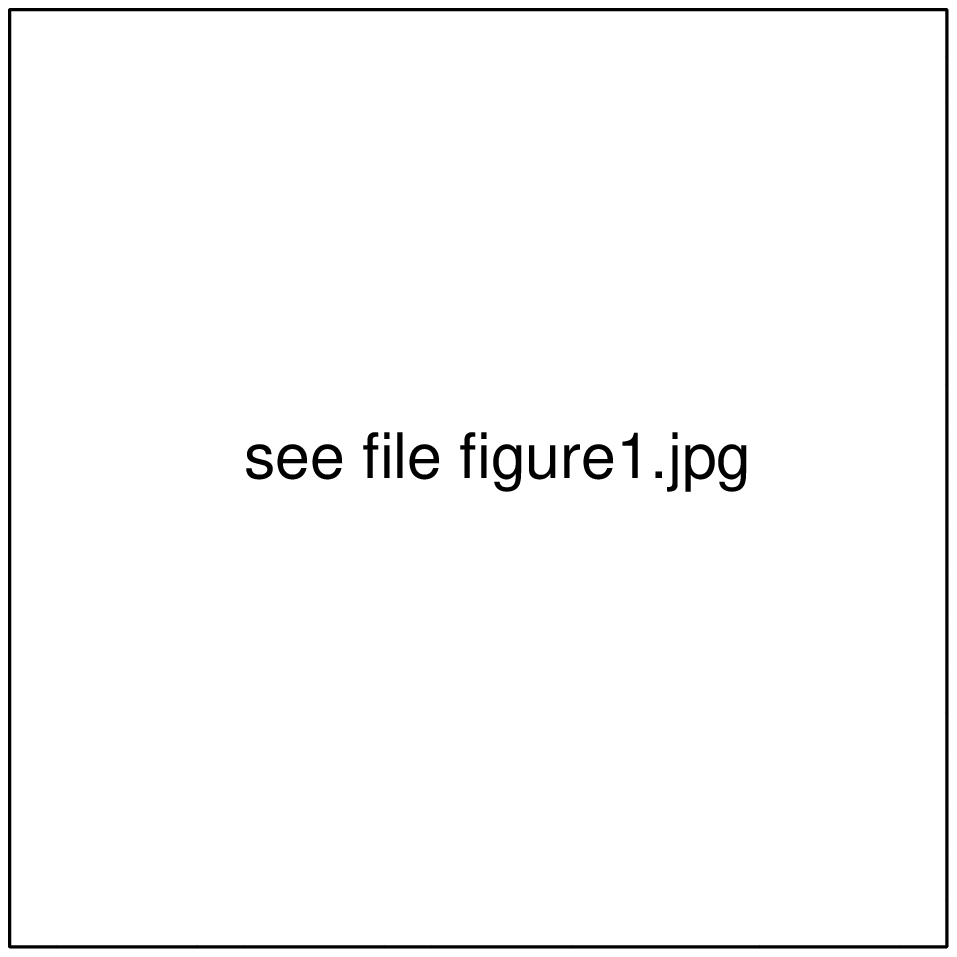}{2.8}{0.45}{-10}{-0}
\figcaption{Observing strategies overlaid on the surface brightness map
  for a typical cluster in our catalog.  The black boxes indicate the
  fields for the four-pointing mosaic program while the white box
  shows the central field of the line broadening program.  Each field
  is 2.9 arcmin on a side, the FOV of {\it Astro-E2}, which corresponds
  to 321 kpc at $z=0.1$.
\label{fig:tiles}}
\vskip0.05in

The thermal plasma emission spectra for this study are compiled using 
the APEC code (Smith et al. 2001) from the XSPEC 11.3.1 suite of spectral analysis
tools (Arnaud 1996).  APEC allows us to include thermal broadening
of the line spectra by setting the APECTHERMAL toggle to `yes'.
Further, XSPEC allows us to fold in the 
anticipated XRS response functions\footnotemark[6]. 
\footnotetext[6]{obtained at
  \url{http://heasarc.gsfc.nasa.gov/docs/astroe/prop\_tools/xrs\_mat.html}} 
We use XSPEC/APEC to write a reference table of spectra in flux units
at 176 temperatures spaced logarithmically to cover our range of
interest.   This spectral table is limited to an energy range appropriate for
studying the Fe K complex.  The final spectra are limited to
the interval 5.9 keV to 6.4 keV.  All spectra are generated
with the cluster at a fiducial redshift of 0.1, so that the FeXXV
K$\alpha$ line is centered at 6.09 keV and the FeXXVI K$\alpha$ line
at 6.33 keV (the hydrogen-like iron line is actually two separate lines
with centers at 6.32 and 6.34 keV as a result of fine structure
(Verner, Verner \& Ferland 1996)). Figure~\ref{fig:thermal} shows the K$\alpha$ line
region of reference spectra at a few relevant temperatures. 

To create a simulated spectrum for a mock observation, we interpolate
on the reference flux table to generate an emission spectrum for each
gas particle of the simulation contained in the {\it Astro-E2} FOV.  There are typically $\sims
1000$ particles in the the FOV, and their contributions are
summed to obtain the complete spectrum.
To convert the resulting flux spectrum into an XRS count spectrum, we
first define an exposure time that requires either 200 or 400 total
counts in the Fe K$\alpha$ line region within the energy range of interest, i.e.,
6.0-6.2 keV.  (The 200 count criterion applies to the four-pointing mosaic study,
while 400 counts are used for the central pointing spectra.)
A single, discrete Poisson realization of each
spectrum is then created, using the
{\it poidev} routine of Press et al. (1992).

\vskip-0.2in
\myputfigure{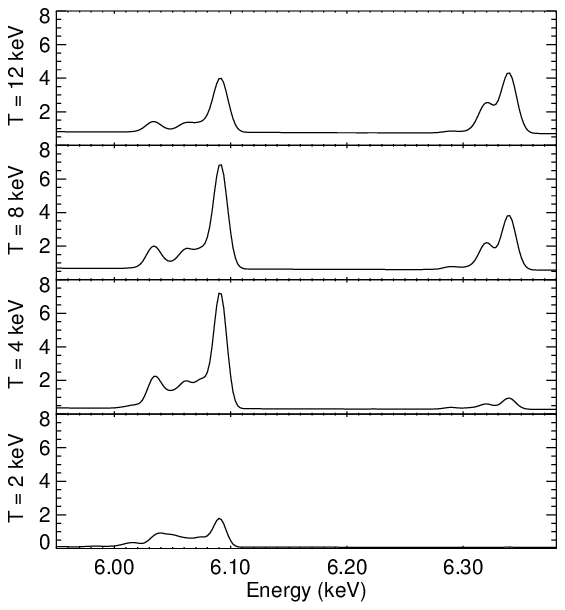}{2.8}{0.45}{-10}{-0}
\figcaption{APEC thermal spectra of 0.4 solar metallicity plasma (arbitrary flux
  units) in a narrow energy range centered on the iron K$\alpha$
  complex at redshift $z \se 0.1$ are shown for typical temperatures
  of plasma in the simulations.  The
  {\it Astro-E2} XRS energy response has been folded in.
\label{fig:thermal}}
\vskip0.1in

\subsection{Additional Assumptions}

The iron abundance of the ICM is not determined by cluster simulations, and
so has to be assumed based on data from observed clusters.
Given the small field of view of the XRS on-board {\it Astro-E2}, we are
probing regions fairly close to the cluster's core ($\lta 0.2\;
r_{200}$).  The iron abundance in cluster cores 
is typically higher than the average over the whole
cluster, due to the frequent presence of central metal
abundance gradients in cold core (``cooling flow'') clusters 
(Ulmer et al. 1987; White et al. 1994), which constitute the majority of clusters. 
Assuming an average central iron abundance of $0.27$ solar
for non-cold core clusters and $0.47$ for cold core systems (DeGrandi
et al. 2004), and recognizing that $70-90\%$ 
of an X-ray flux-limited sample of clusters have cold cores (Edge,
Stewart \& Fabian 1992),  
we obtain an average central iron abundance of $\sim$0.4 relative to the solar photospheric values of 
Anders \& Grevesse (1989).
This is the value assumed throughout this work.
    
One of the main obstacles in measuring velocities of the intracluster
gas with current spectrometers is the inability to accurately
calibrate the temporal and spatial fluctuations of
instrumental gain, the conversion of pulse-height into
photon energy (Dupke \& Bregman 2001b). For the case of CCDs the intrachip
(positional dependent) gain changes are related to the charge transfer
inefficiency, which evolves with time. Often these fluctuations are on
the same order as the velocities one is trying to measure. Although
this is not an issue for the XRS calorimeter, 
small drifts in the temperature of
the detector heat sink can cause global gain variations. The absolute
precision of the gain variations in the XRS is expected to be
$\sims 1-2$~eV ($50-100 \kms$) at 6~keV and will have to be inserted into
the uncertainties of the velocity measurements described here
(Figueroa, personal communication, 2004). These uncertainties were not
incorporated in our evaluations of velocity gradients, but they do not
affect the overall results derived in this work. 
Similarly, for the purpose of this paper we do not fake a background spectrum. 
The main source of background for the XRS is due to energy deposition in the detector by energetic protons.
However, given the small detector size 
and the distribution of the proton energy spectrum, its contribution in the frequencies of interest ($<$ 10 keV)
are expected to be negligible ($<$ 2-5\%) for our mock observations. 

Finally, in this paper we assume that the intracluster plasma is optically thin. 
Gil'fanov, Sunyaev \& Churazov (1987) have pointed out that this may 
not always be true for the Fe K$\alpha$ line. Resonant scattering (the 
absorption and immediate re-emission of an Fe K$\alpha$ line photon by an Fe
ion) can be significant in the cores of clusters. If so, the line 
emission from the cluster core is reduced because of the scattering of photons  
out of the field of view. Early evidence for 
resonant scattering (Molendi et al. 1998) seen in the Perseus cluster has
proven ambiguous, allowing other possible interpretations such as an 
overabundance of Ni due the central SN Ia ejecta dominance (Dupke \& 
Arnaud 2001). {\it XMM} observations of the Perseus cluster have  
not exhibited resonant scattering in the Fe K$\alpha$ line,
and the
absence of
observed scattering has been
used as evidence for the presence of
turbulent (non-thermal) velocities in clusters' cores
(Gastaldello \& Molendi 2004, Churazov et al. 2004).

\vskip-0.2in
\myputfigure{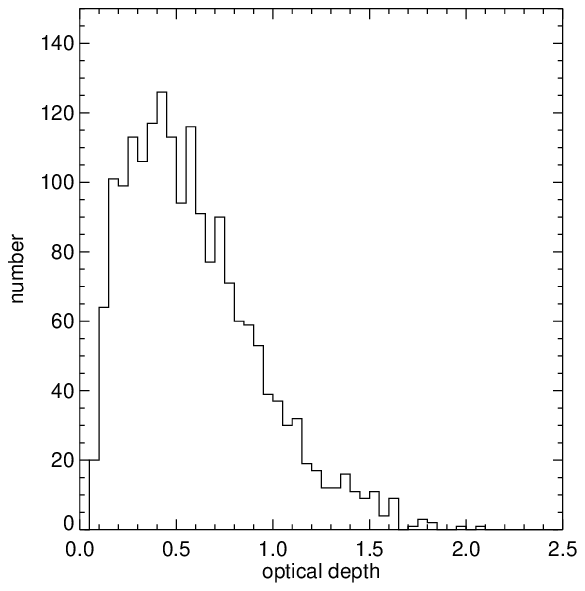}{2.8}{0.45}{-10}{-0}
\figcaption{Histogram of the optical depth for the central
pointing of the 1836 cluster projections used in this
study.
\label{fig:tau}}

To quantify the potential impact of resonant scattering, we compute the optical
depth for the central pointing of each cluster projection used in this study
(see Appendix \ref{sec:appa} for details).
To be conservative, we calculate the optical depth
through the entire cluster.  The results are shown in Figure \ref{fig:tau}.
When the non-thermal velocities seen in the emission 
line are included, the average optical depth for the FeXXV K$\alpha$
line in the FOV of the central pointing is $0.59$, with standard
deviation $0.34$.  The results of Churazov et al. (2004) and Gil'fanov
et al. (1987) show that for $\tau \sim 0.6$, 
the flux and the linewidth near the cluster core will show approximately 
a 15\% reduction.  
Additionally, it is possible that
some deformation of the line profile may occur (Gil'fanov et al. 1987).  
None of these effects would introduce errors in the line centroids we recover.
The reduction in linewidth would affect our line broadening study, but, as our
results will show, a 15\% reduction is not prohibitive.

\subsection{Velocity Gradients Via Multiple Pointings}
\label{sec:velgrad}

We first investigate the LOS velocity structure of the simulated clusters by
creating one set of four mock exposures tiled around the most bound
position of the relevant cluster, as shown by boxes 1-4 in
Figure~\ref{fig:tiles}.  The mock exposures result in four
separate Poisson-noise-added spectra.  We rebin these spectra
to ensure at least 15 counts in each bin using the {\it grppha}
routine from the FTOOLS package\footnotemark[7].
\footnotetext[7]{see
  \url{http://heasarc.gsfc.nasa.gov/docs/software/ftools/ftools\_menu.html}}
We fit the grouped spectra to find the position of the
centroid velocity $v_i$ of the Fe K$\alpha$ line in image $i$.  The
spectra are fit in XSPEC to  
a BAPEC model. 
BAPEC is a thermally broadened APEC that allows for
additional Gaussian velocity broadening under a parameter $\sigma_v$.

The resulting centroid positions are differenced to define velocity
gradients $v_{ij} \se |v_i -v_j|$ among the six distinct pairs of the four
pointings.  Thus, from our 1,836 independent cluster projections we
generate 7,344 spectra which are used to give 11,016 velocity
differences. 

An advantage of working with simulated clusters is our complete
knowledge of the underlying velocity structure of the gas.  In an
isothermal cluster, the redshift of the line center could be recovered
from the density-squared weighted velocity (henceforth
called the EM-weighted velocity) of the emitting gas.  Our 
simulated clusters are not isothermal, but it happens that the Fe K$\alpha$
line flux is not strongly temperature dependent within the temperature 
range expected for hot ICM gas.  The peak flux varies by only a factor
of 2 over a temperature range from 3 keV to 12 keV.  For this reason, an
EM-weighted velocity can be regarded as a predictor of the line center 
that would be recovered from a spectral analysis. 
Therefore, as a point of comparison we also generate the
EM-weighted radial velocity of each {\it Astro-E2} FOV directly
from the simulation.  

To minimize systematic issues, 
we find the velocity of each FOV for
a given cluster projection relative
to the mean value of the four FOV's for that
projection.  We repeat this procedure using the BAPEC
redshift values. 
A histogram of the differences between the
velocities recovered in this way 
from the BAPEC spectral fits and from the EM-weighted averages 
is shown in Figure \ref{fig:allcorr}.  Note that the XRS
instrument is expected to find line centroids with a limiting
resolution of about 2 eV near the Fe K$\alpha$ complex\footnotemark[8],
\footnotetext[8]{see
  \url{ftp://legacy.gsfc.nasa.gov/astroe2/nra\_info/astroe2\_td.pdf}} 
which
corresponds to an expected limiting accuracy of $\sim$100 km/s near the
Fe K$\alpha$ line.  The histogram of differences has a dispersion of
$54 \kms$.  We conclude, therefore, that 200 counts in the line (with
grouping applied)
results in velocity measurement accuracies very comparable to the
expected 2~eV pre-launch instrument response.  This result is
consistent with the findings of Fujita et al. (2005).

\myputfigure{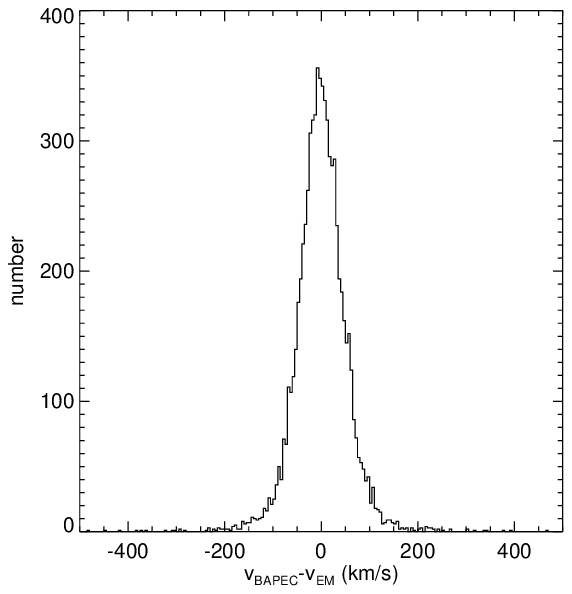}{2.8}{0.45}{-10}{-0}
\figcaption{Histogram of the differences between velocities
  recovered from BAPEC fittings and those generated using
  EM-weighted velocities.
\label{fig:allcorr}}
\vskip0.1in

\subsection{Nonthermal Velocities Via Line Broadening}
\label{sec:zerobroad}

Line broadening is another means of recovering the velocity structure
of the emitting plasma.  As described by Sunyaev et al. (2003) and by
Inogamov \& Sunyaev (2003), the large atomic mass of iron suppresses thermal
line broadening, making the Fe K line width a sensitive probe of bulk
gas motion.  For line broadening studies, only one mock {\it Astro-E2} image
of each cluster centered on the most bound position is used, giving a
total of 1,836 observations.  For this study, our final spectra are
generated by requiring 400 counts in the line region (6.0-6.2 keV).
The velocity broadenings of the spectra are returned as the parameter
$\sigma_v$ of the BAPEC fit.

Once again, we perform a theoretical check on these results.
The EM-weighted velocity dispersion of the particles
within the {\it Astro-E2} field of view is taken to be a predictor of
the best-fit BAPEC $\sigma_v$ parameter.
The differences between these two measures of velocity broadening are
shown in Figure \ref{fig:widcorr}.  The dispersion for the entire data
set of 1,836 broadenings is $55 \kms$.  

The data of Figure \ref{fig:widcorr} show evidence of a systematic
effect.  The mean value of $\sigma_{v}(BAPEC) - \sigma_{v}(EM)$ is
clearly offset from zero.  For the complete sample, it is 
$-26.6 \kms$ (median $-25.4 \kms$).  
This systematic is at least partly explained by the fitting process.  
The spectral
resolution (FWHM) of the XRS is 
about 6.5 eV,
corresponding to a response contribution to the broadening of
$\sigma_{r}$ = 1.5 eV or $\sims 75 \kms$ near the Fe K$\alpha$ line.  This implies that XSPEC
should be unable to recognize physical broadenings significantly less
than $75 \kms$ in our spectra.  
Clusters with EM-weighted broadenings significantly less than 75 km/s are
therefore reported as zero broadening in
the BAPEC fit.  

\myputfigure{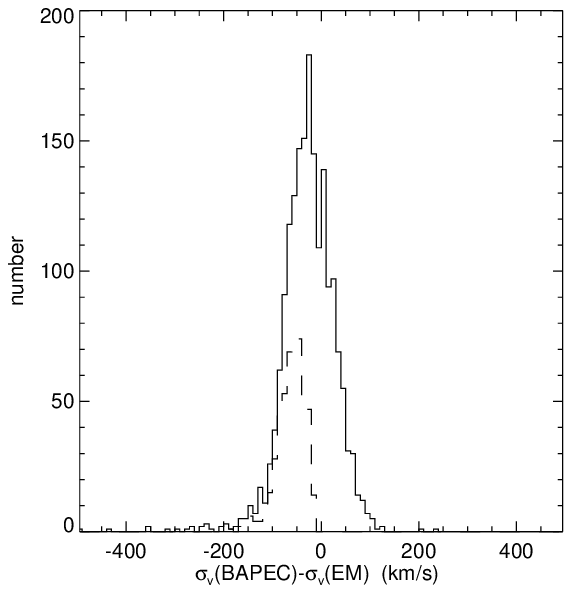}{2.8}{0.45}{-10}{-0}
\figcaption{Histogram of the differences between the velocity broadening
  parameter returned by BAPEC and the EM-weighted velocity
  dispersion.  The dashed line denotes the portion of the sample with BAPEC
uncertainties $> 5 \sigma_v({\rm BAPEC})$ (the ``zero broadening'' sample).
\label{fig:widcorr}}
\vskip0.1in

An investigation of this interpretation can be made using the
BAPEC-reported uncertainty in the broadening value returned.  
If we divide the uncertainty in the broadening parameter by 
the value of the parameter, we expect to obtain a large
number for 
spectra with broadenings that fall below the resolution
of the XRS instrument.  As a test, we construct this ratio
and consider any spectrum with a broadening uncertainty greater
than five times the broadening value to belong to the ``zero-broadening
sample''.  Implementing this definition yields a zero-broadening sample
of 492 clusters 
which has a mean velocity broadening reported by BAPEC of $9.0\times 10^{-3} \kms$
(median $1.8\times 10^{-4} \kms$) and
a standard deviation of $0.025 \kms$.  These numbers are well below the
resolution of the XRS instrument, confirming our intuition that a significant
fraction of the overall cluster sample have physical broadenings unresolvable
by XRS.  The contribution of this zero-broadening sample to the
histogram of Figure \ref{fig:widcorr} is shown as a long-dashed line.
Removing projections with zero reported broadening 
shifts the mean value of $\sigma_{v}(BAPEC) - \sigma_{v}(EM)$ 
to $-13.4 \kms$ (median $-10.4 \kms$) and leaves the dispersion
unchanged.

\section{Results}

\subsection{Velocity Gradients}

Figure~\ref{fig:allshift} shows the cumulative frequency distribution of
velocity differences, normalized to the sound speed in the ICM, that
result from the four-pointing tiling program.   Because the ICM
temperature is determined by the gravitational potential $\phi$ of a 
cluster, the sound speed in the ICM gas is approximately 
\begin{equation}
       c_{s} = \sqrt{\frac{5kT_{ICM}}{3\mu m_{p}}} \lta \sqrt{\phi} 
\label{eq:cs}
\end{equation}
where $\mu m_{p}$ is the average mass per ICM particle.  Since the
bulk velocity structure of the ICM is determined by mergers also
driven by the gravitational potential of the dominant cluster, one
expects infall velocities 
\begin{equation}
        v_{\rm inf} \approx \sqrt{2\phi} .
\label{eq:vinf}
\end{equation}
All explicit dependence on $\phi$, hence on the size of the cluster,
vanishes when infall velocity is scaled by the sound speed.  

\myputfigure{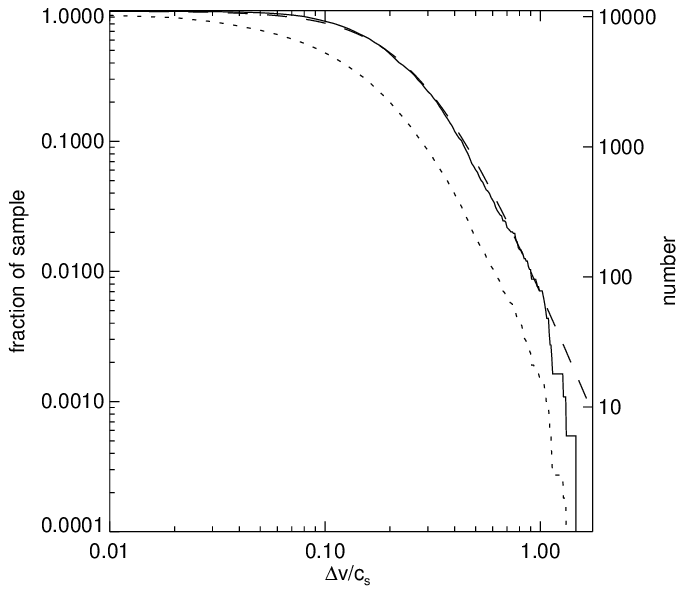}{2.8}{0.45}{-10}{-0}
\figcaption{Cumulative likelihoods of the normalized velocity
gradients $\Delta v/c_s$ are shown for the case of only maximum velocity
differences (solid) and the case of all pointing pairs (dotted).  A
fit to the former is given by equation~(\ref{eq:dvfit}) (dashed).
\label{fig:allshift}}
\vskip0.1in

For each 
simulation, we determine the sound speed from the best-fit spectral
temperature of an isothermal fit to the $0.5-9.5$~keV emission of
plasma within $r_{500}$, the radius within which the mean interior
density is 500 times the critical value.  This radius generally
encloses several thousand of the gas particles making up the simulated cluster.
The $0.5-9.5$~keV spectra
are created by interpolating on a reference table of isothermal
plasma emission spectra to find the 
contribution from each gas particle, and then summing these contributions
to obtain the complete spectrum.   
The procedure is the same as that outlined in \S\ref{sec:mockex} for
creating mock Fe K$\alpha$ line spectra, except that
the table of reference spectra for interpolation is generated using the
MEKAL plasma emission code (Mewe, Kaastra \& Liedahl 1995) and the 
spectral resolution
is 150 eV per bin.  The best-fit spectral temperature is determined
by minimizing the chi-square of the residuals generated by differencing 
the complete
$0.5-9.5$~keV spectrum of the simulation with isothermal plasma spectra
created using MEKAL.

The solid line in Figure~\ref{fig:allshift} shows the cumulative frequency
distribution of the maximum velocity gradient among the six pairs of
each four-pointing mosaic.  The fraction of cluster projections in the 
sample of 1,836 with a maximum normalized velocity difference larger than some
value $x$ is well fit by 
\begin{equation}
  f(\Delta v_{max}/c_{s} > x) \approx \left(1+
  \left(\frac{x}{0.30}\right)^{2}\right)^{-2}. 
\label{eq:dvfit}
\end{equation}
The number of cluster projections in our sample with a maximum
normalized velocity difference larger than half the sound speed for
the cluster is 111, or approximately 6.0\% of the sample.  The total
number of splittings out of the 11,016 measured to be larger than
$0.5c_{s}$ is 200, or approximately 1.8\%. 

The likelihood of high Mach number splittings falls off 
dramatically, as $(\Delta v/c_s)^{-4}$.  The lack of very high
Mach collisions is expected from the arguments given in
equations~(\ref{eq:cs}) and (\ref{eq:vinf}) above.  One anticipates
$v_{\rm inf} \lta$ a few $\times  c_s$ because larger Mach numbers could not
be generated by hierarchical clustering driven by gravity.  Our mock
spectral measurements are consistent with this expectation.  
Note that the quoted value of the sound speed always refers to the
dominant member of a merging pair.  With respect to $c_s$ of the
smaller member, high Mach number collisions are certainly possible.  
But the measured ``broad-beam'' temperature, and sound speed, of the
merging system will always be driven by the larger member, and thus
the data presented in Figure~\ref{fig:allshift} are appropriate for
comparison with observation.  

\myputfigure{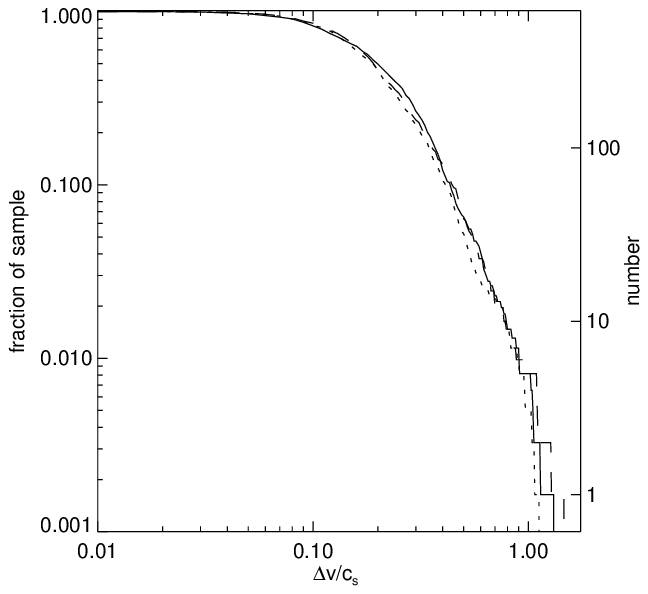}{2.8}{0.45}{-10}{-0}
\figcaption{Cumulative likelihoods of the normalized velocity
gradients $\Delta v/c_s$ are shown for the case of maximum velocity
in three subsamples grouped
according to simulation output redshift.  The short-dashed line is the 
subsample of clusters at $z \ge 0.365$, the long-dashed line is
clusters at $0.365 > z > 0.102$, and the solid line is
clusters at $z \le 0.102$.
\label{fig:newoldshift}}
\vskip0.1in

It is important to recognize that 
even though all clusters have been moved to a
fiducial redshift of $z=0.1$ to construct mock observations, 
we have used clusters recovered from a simulated
sample at several different output
redshifts ranging from $z = 0.540$ to
$z = 0$.  It is therefore natural to ask if the frequency of large
velocity gradients depends on the epoch of observation of the
cluster sample.  To address this question,
we have constructed the cumulative frequency distribution of the maximum
observed normalized velocity gradient for three subsamples grouped by simulation
output redshift.  The results are shown in Figure~\ref{fig:newoldshift}.
The Kolmogorov-Smirnov (K-S) test (see, {\it e.g.}, Press et al. 1992) 
returns a 48.9\% likelihood that the high redshift and middle redshift 
subsamples follow the same distribution, a 44.5\% likelihood that the
low and middle redshift subsamples follow the same distribution, and 
an 8.1\% likelihood that the high and low redshift subsamples follow
the same distribution.

Another question to consider is whether the size of the
{\it Astro-E2} FOV relative to the cluster's size plays a role in the frequency
of large velocity splittings observed.  We address this 
by constructing the cumulative frequency distribution of the
maximum observed normalized velocity gradient for two subsamples 
grouped by the physical size of the cluster (recall that
all clusters are placed at the same fiducial redshift 
for observation, so that physical size corresponds directly to
angular size).  
Clusters with an $r_{200}$ larger than 1.46 Mpc (the median value for our sample)
make up the large cluster subsample, while clusters with $r_{200} < 1.46$ Mpc
form the small cluster subsample.  
The resulting distributions are shown in Figure~\ref{fig:bigsmallshift}.
The K-S test gives a 58.6\% likelihood that both subsamples follow the same
distribution.

\myputfigure{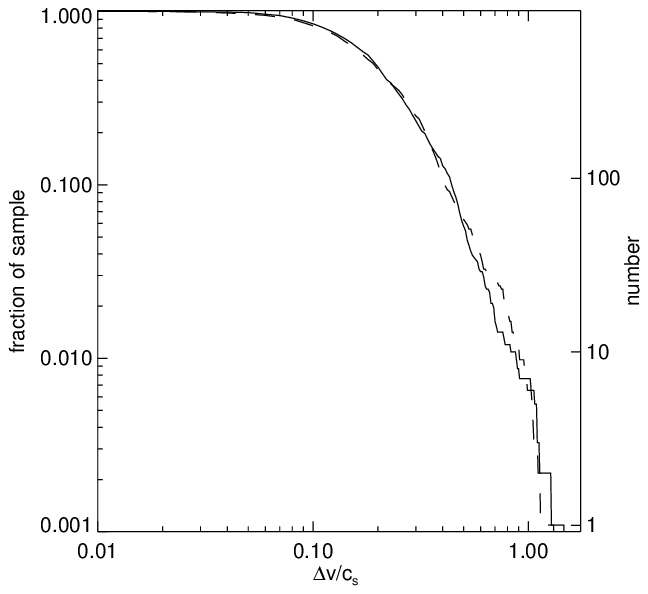}{2.8}{0.45}{-10}{-0}
\figcaption{Cumulative likelihoods of the normalized velocity
gradients $\Delta v/c_s$ are shown for the case of maximum velocity
in two equal subsamples grouped
according to cluster size.  The dashed line is the 
subsample of clusters with  
$r_{200} < 1.46$ Mpc, the solid
line is clusters with $r_{200} > 1.46$ Mpc.
\label{fig:bigsmallshift}}
\vskip0.1in

\begin{figure*}
\begin{center}
\includegraphics[width=6.25in]{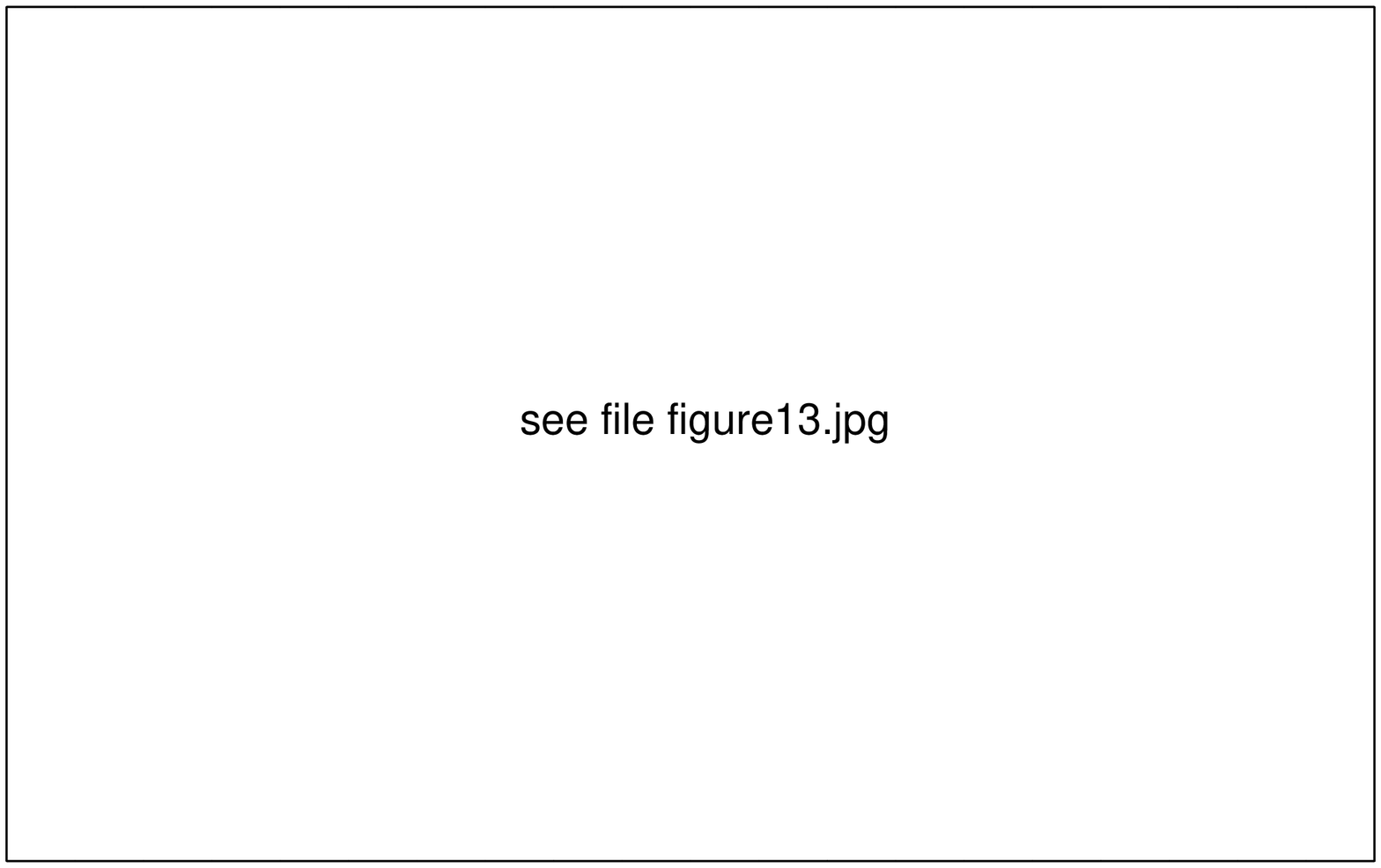}
\end{center}
\caption{Temperature maps (keV) for cluster 2 at output 
redshifts 0.290
(left), 0.222 (middle), and 0.160 (right) along the $x$ (top) and
  $y$ (bottom) axes.} 
\label{fig:spltemp}
\end{figure*}

For purposes of illustration, we 
highlight the cluster which exhibits the maximum velocity
gradient: cluster number 2 at output redshift 0.222
in $y$-axis
projection.  This projection yields a maximum normalized
velocity gradient of 1.47
$c_{s}$.  Data relevant to that cluster are compiled in Figures
\ref{fig:splspec}-\ref{fig:spltemp}.  Figure~\ref{fig:splspec} shows
Poisson realizations of the four-point mosaic spectra created
using the procedure outlined in \S\ref{sec:mockex} and grouped
using the procedure explained in \S\ref{sec:velgrad}.  These spectra
show that the top two quadrants of our observing pattern
are blueshifted relative to the
bottom two.  This accurately represents the underlying motion
of the cluster, as can be seen from Figure \ref{fig:persplspec} which
shows the perfect
flux spectra generated from the simulation overlaid with the BAPEC
fits to the spectra of Figure \ref{fig:splspec}.  

\begin{figure*}
\begin{center}
\includegraphics[width=6.25in]{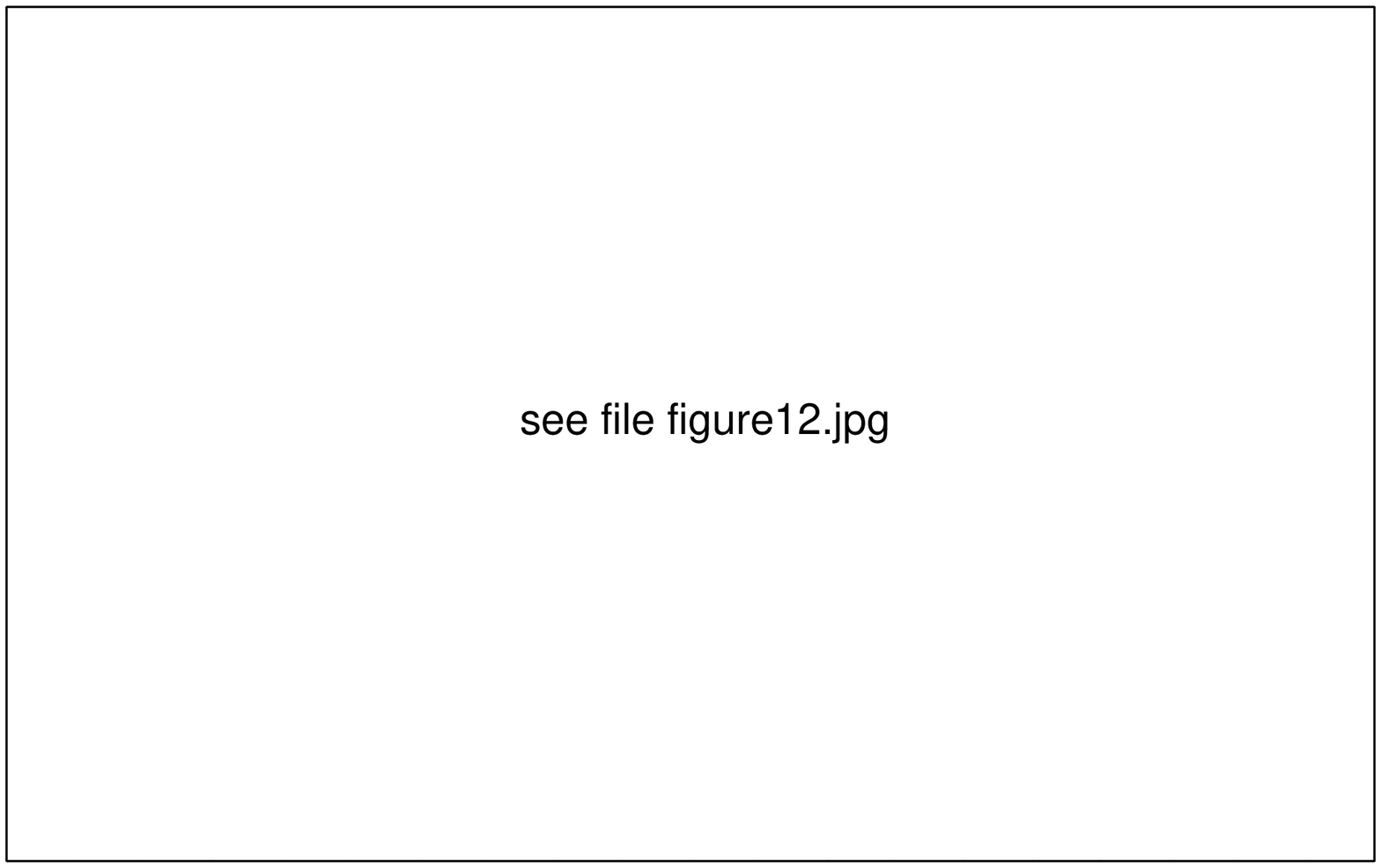}
\end{center}
\caption{Maps show the log of the unabsorbed soft (0.5-2 keV) X-ray surface brightness
  (arbitrary units) for cluster 2 at output redshifts 0.290 (left), 0.222 (middle)
  and 0.160 (right), projected along the $y$ (top) and $z$ (bottom) axes
  of the simulation volume.  
  The width of the image
  box is scaled to $2\: r_{200}$ at each epoch.
  The mock {\it Astro-E2}
  fields of view are overlaid on the frame that displays a velocity
  gradient $\Delta v \se 1.47 c_s$. Note that although this map 
  places the observer along the positive y-axis, the spectra were 
  generated with positive y-axis velocity yielding redshift and negative
  y-axis velocity giving blueshift.}
\label{fig:splsurf}
\end{figure*}

\myputfigure{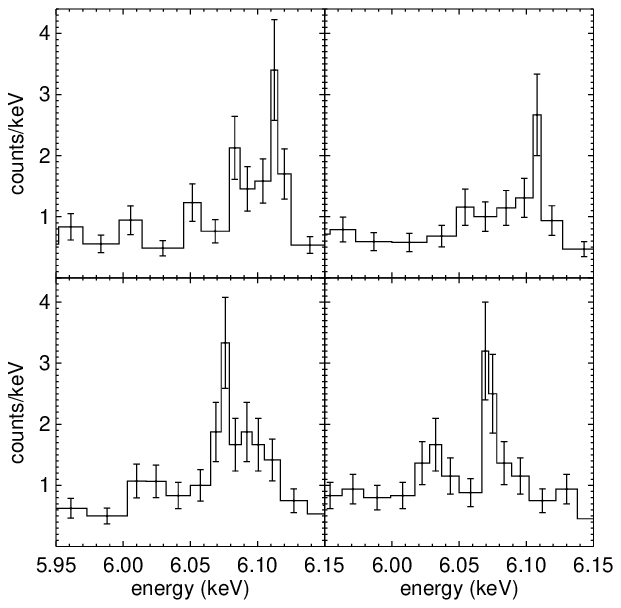}{2.8}{0.45}{-10}{-0}
\figcaption{The integrated and grouped Poisson-noised spectra used in the
velocity gradient study for cluster 2 at $z$ = 0.222, viewed along
the y-axis.  The spectra are generated with the cluster at a fiducial
redshift of 0.1.  Each panel displays the spectrum of the corresponding
four-point mosaic tile pattern shown in Figure~\ref{fig:splsurf}.
\label{fig:splspec}}
\vskip0.1in

\myputfigure{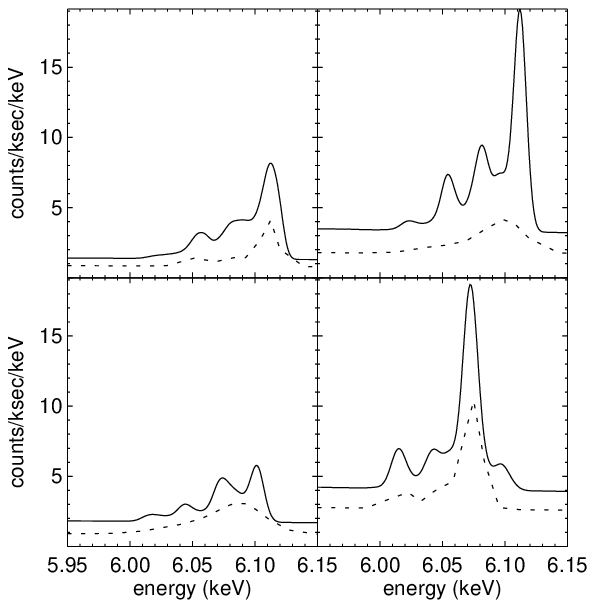}{2.8}{0.45}{-10}{-0}
\figcaption{The flux spectra generated using the APEC thermal
plasma code (solid) for cluster 2 at $z$ = 0.222 are shown along with the
BAPEC fits to the Poisson spectra of Figure \ref{fig:splspec}
(dashed).  The spectra are generated with the cluster at a fiducial
redshift of 0.1.
\label{fig:persplspec}}
\vskip0.1in

The advantage of
using simulations is that we can examine the underlying
cause of this observed velocity structure.  Figure~\ref{fig:splvel}
shows the density-weighted, LOS velocity map of the gas in cluster
2 in a region $2 r_{200}$ on a side.  A strong vertical gradient
in projected velocity is apparent.
The lower
row of images in Figures \ref{fig:splsurf} and \ref{fig:spltemp} 
shows the z-axis (orthogonal view) surface brightness and
temperature maps 
of cluster 2 at output redshifts $z$ = 0.290, 0.222 and 0.160.
 With this additional
information, we can see that cluster 2 is extremely active.  
In this case it appears that the redshifted quadrants (the lower
half of the mosaic) represent the primary cluster itself, which has a peculiar
velocity away from the observer.  The blueshifted quadrants appear
to contain a merging subcluster which is moving through the core
in the negative y-direction.  This subcluster can be traced in
the z-axis projections starting in the upper left corner and
moving toward the lower
right over the course of the outputs presented in Figure \ref{fig:splsurf}.

\myputfigure{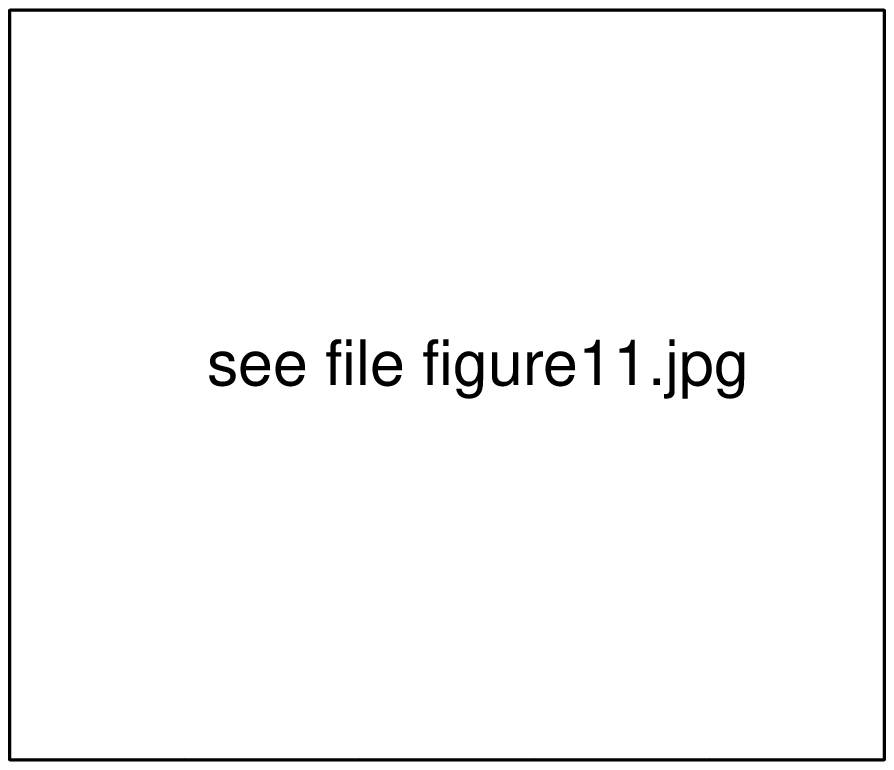}{2.8}{0.45}{-10}{-0}
\figcaption{The line-of-sight velocity map for cluster 2, $z$ = 0.222
  projected along the y-axis is shown along with the mock {\it Astro-E2}
  fields of view.  The scale key is in $\kms$, with positive
  velocities away from the observer.  The map size is $2 r_{200}$ on a side.
\label{fig:splvel}}


\subsection{Line Broadening}
\label{sec:broadresults}

To evaluate the significance of the observed line broadenings, we
normalize $\sigma_v$ recovered from BAPEC by the expected 
thermal velocity dispersion ($\sigma_{th} = \sqrt{kT/m_{\rm Fe}}$).  The
resulting velocity dispersion distribution is presented in
Figure \ref{fig:widhist}.  The distribution ignores $\sim 27\%$ of the fits
that return zero for the velocity broadening parameter (see \S \ref{sec:zerobroad}).
The distribution of non-zero values is fit reasonably well by
\begin{equation}
\label{eq:brdfit}
        f(\sigma_{v}/\sigma_{th} > x) \approx
        0.73\left(1+\left(\frac{x}{1.5}\right)^{3}\right)^{-4/3}. 
\end{equation}
Out of the whole sample, 902 cluster projections, or $49\%$,
yield a BAPEC velocity dispersion larger than the thermal broadening
of the cluster.  As in the gradient case, the probability drops as
$(\sigma_v/\sigma_{th})^{-4}$.  

As we did for the velocity gradient study, we can also consider the impact of
the epoch of observation and of the relative size of the {\it Astro-E2}
FOV on the results of Figure~\ref{fig:widhist}.  To determine if our
results are
dependent on the epoch of observation, we again construct three subsamples 
grouped by simulation output redshift
and generate the cumulative frequency of normalized velocity broadenings
for each subsample.  The results are shown in Figure~\ref{fig:newoldwid}.
The K-S test returns an 89.5\% likelihood that the high redshift and middle
redshift subsamples follow the same distribution, a 29.3\% likelihood that
the middle and low redshift subsamples follow the same distribution, and
a 26.2\% likelihood that the high and low redshift subsamples follow the
same distribution.

\myputfigure{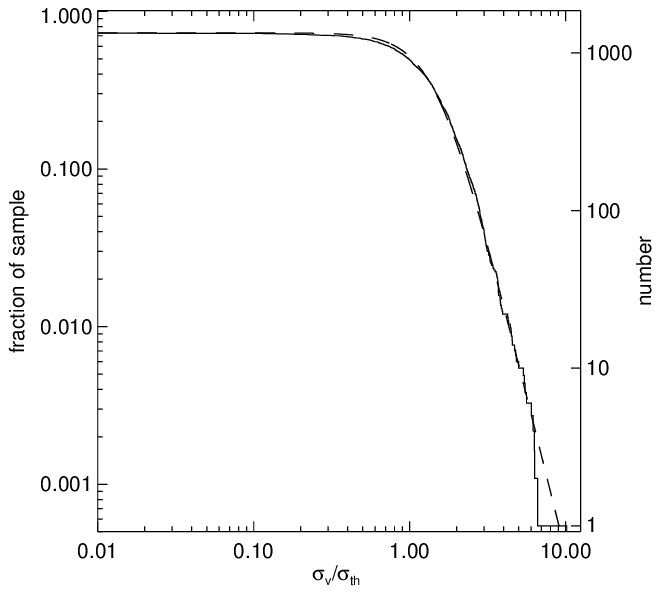}{2.8}{0.45}{-10}{-0}
\figcaption{Fraction of sample with normalized velocity broadening
greater than the x-axis value.  The dashed line represents the
fit given in equation (\ref{eq:brdfit}).
\label{fig:widhist}}
\vskip0.2in

\myputfigure{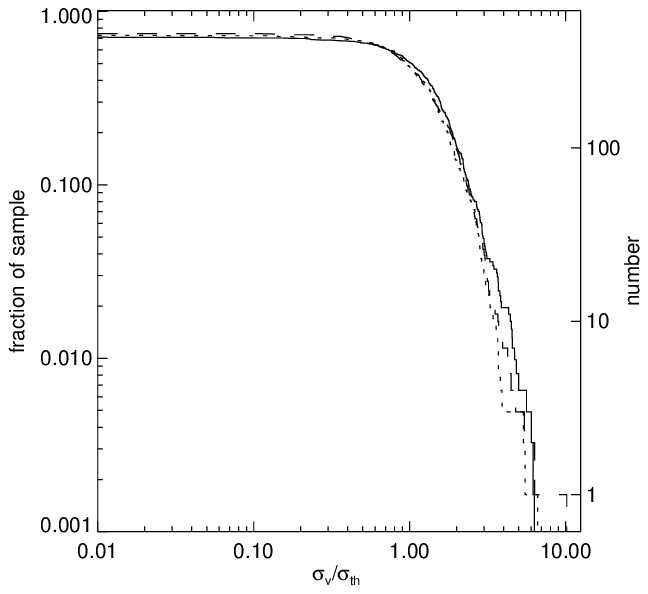}{2.8}{0.45}{-10}{-0}
\figcaption{Cumulative likelihoods of the normalized velocity broadening
for three subsamples grouped according 
to simulation output redshift.  The short-dashed line is the
subsample of clusters at $z \ge 0.365$, the long-dashed line is
clusters at $0.365 > z > 0.102$, and the solid line is
clusters at $z \le 0.102$.
\label{fig:newoldwid}}
\vskip0.2in

 To evaluate the significance
of the size of the {\it Astro-E2} FOV on the observed clusters, we create two subsamples
grouped by the physical size of the cluster.  The resulting cumulative
frequency distributions are shown in Figure~\ref{fig:bigsmallwid}.
For these distributions a K-S test is inappropriate because the result
is dominated by the different sizes of the ``zero-broadening''
component of each subsample.  Recall from \S \ref{sec:zerobroad} that
clusters with velocity broadenings less than about $50 \kms$ are expected
to be unresolved by the XRS instrument.  Clusters with smaller physical
sizes are expected to have smaller physical broadenings.  This fact implies that
the small cluster subsample should exhibit a larger fraction of broadenings below 
the resolution
of the XRS instrument than the large cluster subsample does.  
This effect is seen in the low-broadening limiting
values of the distributions of Figure \ref{fig:bigsmallwid}.  This difference
dominates the K-S test, resulting in a likelihood of $1.5\times 10^{-7}$ that
the large cluster and small cluster subsamples follow the same distribution.
If we perform the K-S test without including differences in
the region $\sigma_{v}/\sigma_{th} < 2$
we find a 98\% likelihood that the distributions are the same, implying
that the apparent divergence at large broadenings is statistically insignificant.  

\myputfigure{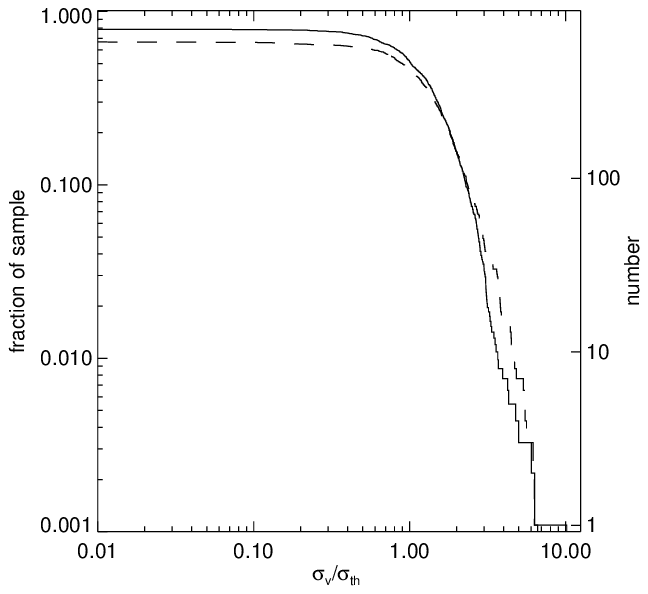}{2.8}{0.45}{-10}{-0}
\figcaption{Cumulative likelihoods of the normalized velocity
broadening 
for two equal subsamples grouped
according to cluster size.  The dashed line is the
subsample of clusters with $r_{200} < 1.46$ Mpc,
the solid
line is clusters with $r_{200} > 1.46$ Mpc.
\label{fig:bigsmallwid}}
\vskip0.2in

\begin{figure*}
\begin{center}
\includegraphics[width=6.25in]{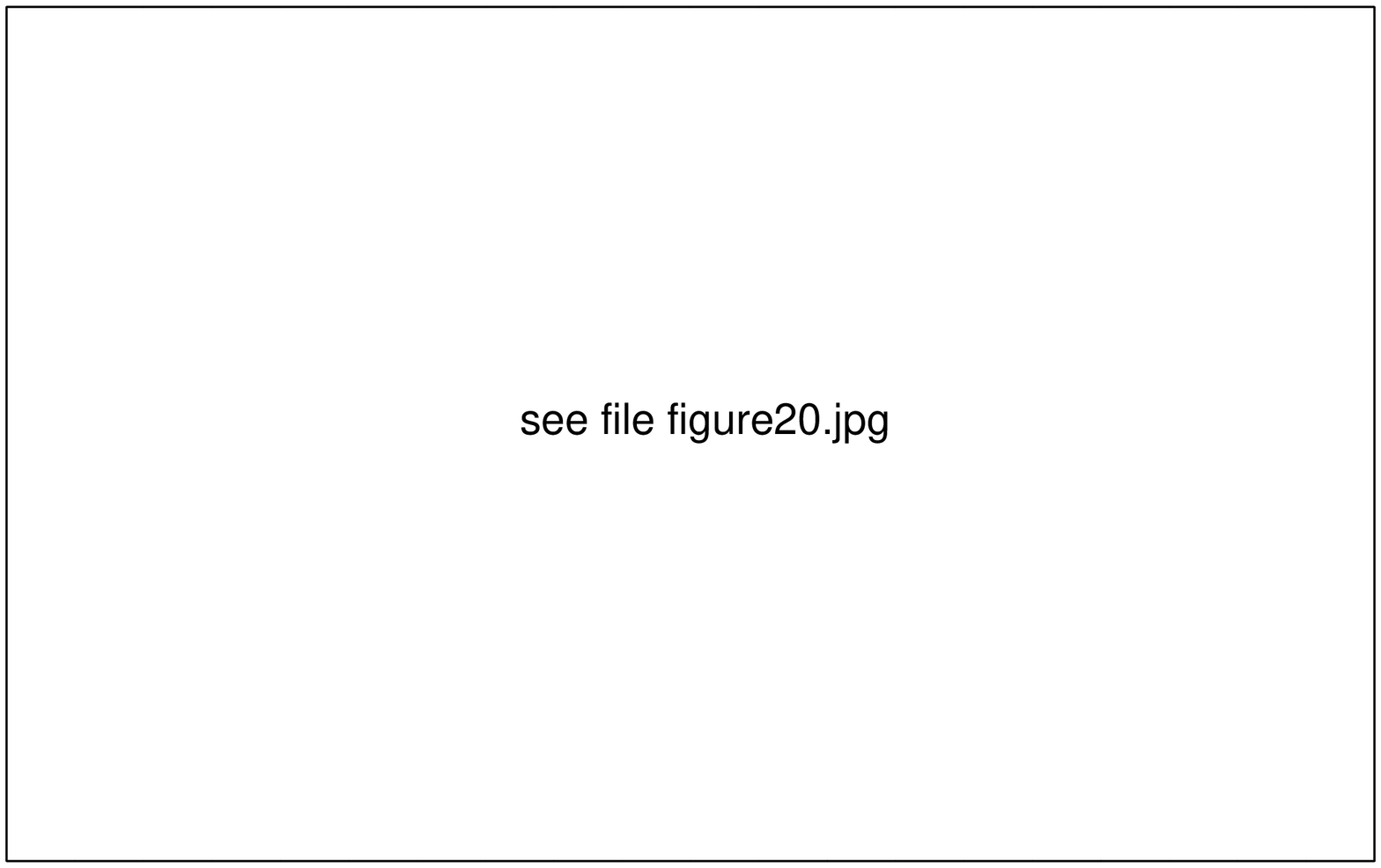}
\end{center}
\caption{Temperature maps (keV) for cluster 47 at output redshifts 0.290 (left), 0.222 (middle) and 0.160 (right) along the x (top) and y (bottom) axes.}
\label{fig:broadtemp}
\end{figure*}

Once again, we select one cluster projection to highlight as
a case study.  We select the cluster which exhibited the
most significant line broadening:
cluster 47 at output redshift 0.222 viewed along the x-axis.  The data
relevant to this cluster is compiled in Figures \ref{fig:broadspec}-\ref{fig:broadtemp}.
Again it is apparent that the cluster is undergoing a strong
merger
at this epoch.  In this case, however, the merging cluster is almost
directly behind cluster 47 (rather than slightly off-center as in the
case of cluster 2 output redshift 0.222 shown above).  This on-axis configuration
means that
all {\it Astro-E2} spectra in the four-pointing mosaic contain emission from
the merging cluster, resulting in nearly identical
(two-component) spectra.  For this reason, each image in the
multiple-pointing study gives approximately the same redshift when fit
with a single-component plasma emission model,
and differencing multiple
images yields small velocity gradients.
In fact, while
this projection gives a
velocity dispersion more than ten times the thermal
velocity broadening, its maximum velocity
gradient in the four-pointing mosaic is
$0.63 c_{s}$.
This
illustrates the complementarity of line broadening and multiple-imaging
studies for detecting dynamically active clusters.

\myputfigure{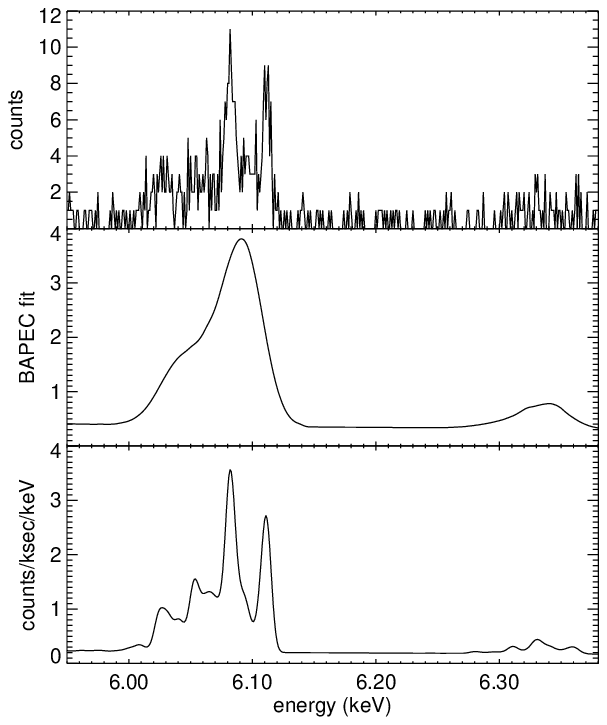}{2.8}{0.45}{-10}{-0}
\figcaption{From top to bottom: the integrated spectrum,
BAPEC fit, and actual flux
spectrum for cluster 47, output at redshift 0.222, taken along the x-axis.
The spectra are generated with the cluster at a fiducial redshift of 0.1.
\label{fig:broadspec}}
\vskip0.1in

\myputfigure{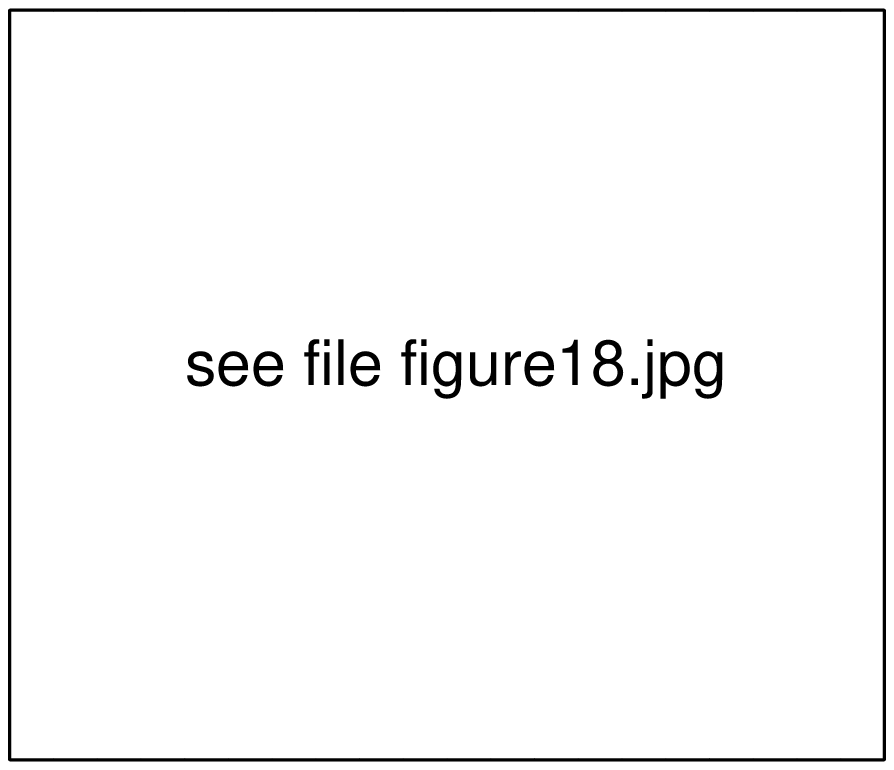}{2.8}{0.45}{-10}{-0}
\figcaption{Radial velocity map for cluster 47, output redshift 0.222 
projected along the x-axis in the same format as
Figure~\ref{fig:splvel}.
\label{fig:broadvel}}
\vskip0.1in

\begin{figure*}
\begin{center}
\includegraphics[width=6.25in]{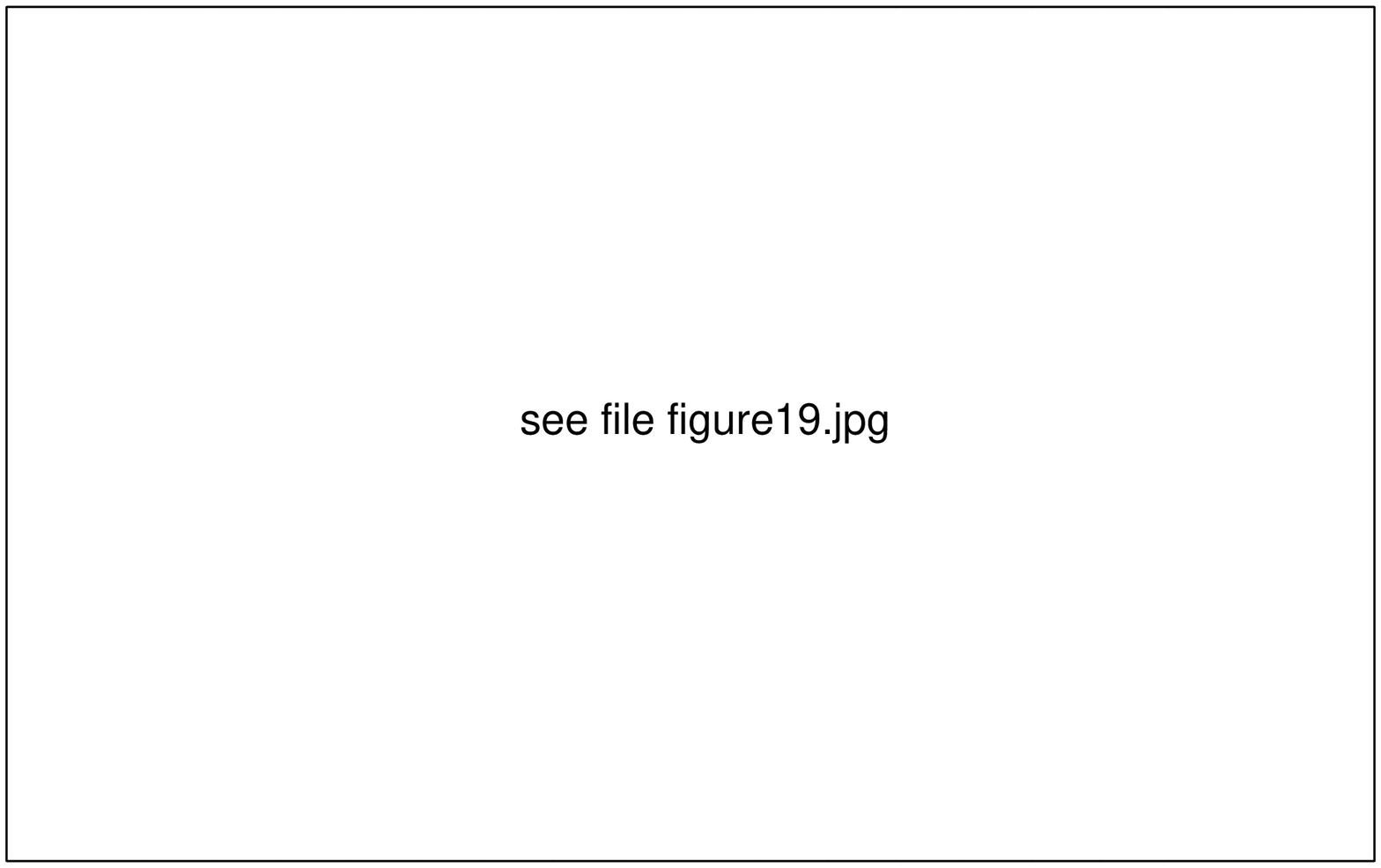}
\end{center}
\caption{Maps show the log of the unabsorbed soft (0.5-2 keV) X-ray surface brightness
  (arbitrary units) for cluster 47 at output redshifts 0.290 (left), 0.222 (middle)
  and 0.160 (right), projected along the $x$ (top) and $y$ (bottom) axes
  of the simulation volume.  
  The width of the image
  box is scaled to $2 r_{200}$ at each epoch.
  The mock {\it Astro-E2}
  fields of view are overlaid on the frame that displays the largest
  velocity broadening of the sample.  Note that although this map 
  places the observer along the positive x-axis, the spectra were
  generated with positive x-axis velocity yielding redshift and negative
  x-axis velocity giving blueshift.
}
\label{fig:broadsurf}
\end{figure*}

\section{Conclusions}

We use a large statistical sample of mock {\it Astro-E2} observations to
investigate the expected frequency of detection of bulk motion of the
ICM using the Fe XXV K$\alpha$ line.  
We analyze differences
in mean velocities  
among adjoining pointings as well as velocity broadenings in a single
central pointing of clusters developed in a 
$\Lambda$CDM model and scaled to an observed redshift $z
\se 0.1$.  The mean velocities and broadenings are recovered
from BAPEC fits to mock {\it Astro-E2} spectra assuming
200 counts
and 400 counts in the line region, respectively.

From the central pointings, we find that roughly one-half of randomly
selected clusters display velocity broadening in the Fe K$\alpha$ line
at least as large as the thermal width, and more than
$15\%$ will be broadened at
a level greater than $2\sigma_{th}$.   In the four-point
mosaic case, $6\%$ of clusters show a maximum velocity 
difference that is transonic ($\Delta v/c_s \ge 0.5$) while only $0.7\%$
will be supersonic ($\Delta v/c_s \ge 1$).

These
fractions were obtained without any attempt to select for interesting
clusters or to optimally align observations on the basis of
morphology.  Further, our sample is biased toward cooler (smaller)
clusters.  We have attempted to remove the effects of this bias by
normalization, but the physical cut represented by the $\sim 75 \kms$
resolution of the XRS instrument
cannot be normalized away.   We can only
assume that an observational survey, conducted to optimize results by
selecting hot clusters with bimodal galaxy velocity distributions 
would obtain a sample with an even larger
fraction of dynamically interesting clusters.  

In extreme cases, however, morphological clues from X-ray observations
can be misleading.  We have illustrated one example in which the viewing
angle is nearly aligned with the infall
direction of a merging satellite. Such cases possess relatively
simple X-ray morphologies but show large velocity effects in the Fe
K$\alpha$ line.  

This calibration exercise bears out the expectation that 
{\it Astro-E2} will be an important instrument for
detecting bulk motion of ICM gas.  With reasonable counting statistics,
the true emission-weighted velocity within the {\it Astro-E2} FOV can be
obtained to within $\sim 50 \kms$, assuming the pre-launch gain and
response estimates are 
accurate and that background is negligible.

\acknowledgments

We are very grateful to K. Arnaud for providing valuable input on spectral
analysis tools.  We also thank E. Figueroa for providing helpful 
information about {\it Astro-E2} gain.  AP thanks J. Bialek
for assistance with the manipulation of the VCE dataset and maps.

This work was supported by NASA through {\it Chandra} Theory grants
TM3-4009X and TM4-5008X and by the NSF through ITR grant ACI-0121671.
RAD also acknowledges support from NASA Grants NAG 5-3247 and {\it 
Chandra} GO3-4162X.  AEE acknowledges support from a JSPS Invitation 
Fellowship to Tokyo University and thanks Y. Suto for hospitality.

FITS format spectra and maps similar to those presented in
this paper are available on the VCE website 
for selected cluster projections used in this study, including all those
which exhibited splittings greater 
than $0.75 c_{s}$ or broadenings larger than $2.5\sigma_{th}$.

\appendix

\section{Computing Optical Depth}
\label{sec:appa}

The optical depth for line emission obeys (Spitzer 1978):
\begin{equation}
  \int \tau(\nu) d\nu = f_{jk}N_{j}\frac{\pi e^{2}}{m_{e}c}\left(1-\frac{b_{k}}{b_{j}}e^{-h\nu_{jk}/kT}\right)
\end{equation}
where $j$ and $k$ label atomic states, $\nu_{jk}$ is the central frequency of the line radiation emitted in
a transition from $k \rightarrow j$, $f_{jk}$ is the upward oscillator stregth for the transition, 
and $N_{j}$ is the column density of atoms/ions in state $j$.  The quantities $b_{k}$ and $b_{j}$ measure
the particle densities $n_{k}$ and $n_{j}$ relative to the values they would have in thermodynamic equilibrium (see
Spitzer (1978) for details).  At ICM densities, we expect $b_{k} = 0$ for the Fe K$\alpha$ line
(Dere et al. 1997; Young et al. 2003).  
The oscillator strength for the K$\alpha$ line of Fe XXV is 0.798 (Verner et al. 1996).  In cgs units, then:
\begin{equation}
        \int \tau(\nu) d\nu \approx 0.0212 N_{j} 
\end{equation}
where $N_{j}$ is the column density (in ${\rm cm}^{-2}$) of Fe XXV in the ground state.  

We next simplify things by defining an average optical depth for the line:
\begin{equation} 
        2\sqrt{2}\:\bar{\tau} \nu_{jk}\frac{\sigma_{jk}}{c} \equiv \int \tau(\nu) d\nu
\end{equation}
where we define a Gaussian sigma parameterizing the width of the Fe K$\alpha$ line.
In the terminology of
this paper, we use the definition:
\begin{equation}
        \sigma_{jk} \equiv \sqrt{\sigma_{th}^{2} + \sigma_{v}^{2}}
\end{equation}
(note that, in keeping with the rest of this work,
we have chosen a $\sigma$ in velocity units so that the combination
$\sigma/c$ is unitless).
We have computed $\sigma_{jk}$ for each cluster as a part of our line broadening study (see
\S\ref{sec:broadresults}).  
We can now write (again, in cgs units):
\begin{equation}
   \bar{\tau} \approx \frac{4.6\times 10^{-21} c N_{j}}{\sqrt{\sigma_{th}^{2} + \sigma_{v}^{2}}}. 
\end{equation}
The
final quantity remaining to compute is $N_{j}$.  

Assuming an iron abundance of 0.4 solar, we can write:
\begin{equation}
        N_{Fe} \approx 2\times 10^{-5} N_{H}.
\end{equation}
Not all the iron will be in the FeXXV ionization state, however.  At maximum,  
78\% of the Fe will be Helium-like.  This fraction
has a non-negligible temperature dependence over the range of temperatures
expected in the ICM.  The peak occurs at a temperature of 3 keV.  For temperatures
below 1 keV or above 8 keV the fraction has fallen below 33\%.  
For this reason, we use our knowledge of the underlying simulation to approximate
the optical depth for resonant scattering in our sample of clusters.  We
define a function $f(T)$ using a linear interpolation on Table 4 of Arnaud \&
Raymond (1992).  We then assign the $i$th gas parcel within the FOV a column density
for resonant scattering given by:
\begin{equation}
       N_{FeXXV}^{i} = 2\times 10^{-5} f(T_{i}) N_{H}^{i}.
\end{equation}
We can then sum up the column densities in the FOV to find the average optical depth
in the FOV:
\begin{equation}
        \bar{\tau}_{FOV} = 9.2\times 10^{-26} \sum_{i \in FOV}\frac{c f(T_{i}) N_{H}^{i}}{\sqrt{\sigma_{th}^{2} + \sigma_{v}^{2}}}.
\end{equation}


\begin{references}

\bref
Anders, E. \& Grevesse, N. 1989, Geochim. Cosmochim. Acta, 53, 197

\bref
Arnaud, K.A. 1996, Astronomical Data Analysis Software and Systems V, eds. Jacoby, G. and
Barnes, J., ASP Conf. Series volume 101 [\url{http://heasarc.gsfc.nasa.gov/\\docs/xanadu/xspec/index.html}]

\bref
Arnaud, M \& Raymond, J. 1992, ApJ, 398, 394

\bref 
Bennet, C.L., et al. 2003, ApJS, 148, 1

\bref
Bialek, J.J., Evrard, A.E. \& Mohr, J.J. 2001, ApJ, 555, 597

\bref
Bialek, J.J., Evrard, A.E. \& Mohr, J.J. 2002, ApJ, 587, L9

\bref
Bialek, J.J., Evrard, A.E. \& Mohr, J.J. 2005, in preparation

\bref
Buote, D.A. \& Tsai, J.C. 1996, ApJ, 458, 27

\bref
Churazov, E., Forman, W., Jones, C., Sunyaev, R. \& B\"ohringer, H. 2004, MNRAS, 347, 29

\bref
De Grandi, S., Ettori, S., Longhetti \& M. Molendi, S. 2004, A\&A, 419, 7

\bref
Dere, et al. 1997, AASS, 125, 149

\bref
Dupke, R.A. \& Arnaud, K.A. 2001, ApJ, 548, 141

\bref
Dupke, R.A. \& Bregman, J.N. 2001a, ApJ, 547, 705

\bref
Dupke, R.A. \& Bregman, J.N. 2001b, ApJ, 562, 266

\bref
Edge, A.C., Stewart, G.C. \& Fabian, A.C. 1992, MNRAS, 258, 177

\bref
Evrard, A.E. 1988, MNRAS, 235, 911

\bref
Frenk, C.S., et al. 1999, ApJ, 525, 554

\bref
Fujita, Y., Matsumoto, T., Wada, K. \& Furusho, T. 2005, ApJ, 619, L139

\bref
Furusho, T., Mitsuda, K., Yamasaki, N., Fujimoto, R., \& Ohashi, T. 2004, 
astro-ph/0401490

\bref
Gastaldello, F. \& Molendi, S. 2004, ApJ, 600, 670

\bref
Gil'fanov, M.R., Sunyaev, R.A. \& Churazov, E.M. 1987, Pis'ma Astron. Zh. 13, 7

\bref
Inogamov, N.A. \& Sunyaev, R.A. 2003, Astron. Lett., 29, 791

\bref
Jones, C. \& Forman, W. 1984, ApJ, 276, 38

\bref
Mewe, R., Kaastra, J.S. \& Liedahl, D.A. 1995, Legacy, 6, 16

\bref
Mohr, J.J., Fabricant, D.G. \& Geller, M.J. 1993, ApJ, 413, 492

\bref
Molendi, S., Matt, G., Antonelli, L.A., Fiore, F., Fusco-Femiano, R., Kaastra, J., Maccarone, C. \& Perola, C. 1998, ApJ 499, 608

\bref
Press, W.H., Teukolsky, S.A., Vetterling, W.T. \& Flannery, B.P.
 1992, {\it Numerical Recipies in Fortran, 2nd ed.}, Cambridge University Press

\bref
Smith, R.K., Brickhouse, N.S., Liedahl, D.A. \& Raymond, J.C. 2001, ApJ 556, L91
[\url{http://cxc.harvard.edu/atomdb//index.html}]

\bref
Spitzer, L. 1978, {\it Physical Processes in the Interstellar Medium}, (Wiley, NY) 

\bref
Sunyaev, R.A., Norman, M.L. \& Bryan, G.L. 2003, Astron. Lett., 29, 783

\bref
Thomas, P.A. \& Couchman, H.M.P 1992, MNRAS, 257, 11

\bref
Ulmer, M.P., Cruddace, R.G., Fritz, G.G., Snyder, W.A. \& Fenimore, E.E. 1987, ApJ, 319, 118

\bref
Verner,~D.A., Verner,~E.M. \& Ferland,~G.J. 1996, Atomic Data and Nuclear Tables, 64, 1

\bref
White, R.E. III, Day, C.S.R., Hatsukade, I. \& Hughes, J.P. 1994, ApJ, 433, 583

\bref
Young, et al. 2003, ApJSS, 144, 135




\end{references}
\end{document}